\documentclass[conference]{IEEEtran}
\IEEEoverridecommandlockouts
\usepackage{cite}
\usepackage{amsmath,amssymb,amsfonts}
\usepackage{algorithmic}
\usepackage{graphicx}
\usepackage{textcomp}
\usepackage{xcolor}
\usepackage{placeins}
\usepackage{float}
\usepackage{adjustbox}
\usepackage{subcaption}

\usepackage{listings}
\usepackage{xcolor}

\usepackage{array}
\usepackage{makecell}

\newcolumntype{C}[1]{>{\centering\arraybackslash}m{#1}}

\def\BibTeX{{\rm B\kern-.05em{\sc i\kern-.025em b}\kern-.08em
    T\kern-.1667em\lower.7ex\hbox{E}\kern-.125emX}}
\begin{document}

\title{A Comprehensive Overview of GPU Accelerated Databases\\
}

\author{
\IEEEauthorblockN{Harshit Sharma$^*$}
\IEEEauthorblockA{\textit{Department of Computer Engineering} \\
\textit{University of Wisconsin-Madison}\\
Madison, Wisconsin, United States \\
hsharma27@wisc.edu}
\and
\IEEEauthorblockN{Anmol Sharma$^*$}
\IEEEauthorblockA{\textit{Department of Computer Engineering} \\
\textit{University of Wisconsin-Madison}\\
Madison, Wisconsin, United States \\
sharma265@wisc.edu}
}

\maketitle

\def\thefootnote{*}\footnotetext{These authors contributed equally to this work}

\begin{abstract}
Over the past decade, the landscape of data analytics has seen a notable shift towards heterogeneous architectures, particularly the integration of GPUs to enhance overall performance. In the realm of in-memory analytics, which often grapples with memory bandwidth constraints, the adoption of GPUs has proven advantageous, thanks to their superior bandwidth capabilities. The parallel processing prowess of GPUs stands out, providing exceptional efficiency for data-intensive workloads and outpacing traditional CPUs in terms of data processing speed. While GPU databases capitalize on these strengths, there remains a scarcity of comparative studies across different GPU systems. In light of this emerging interest in GPU databases for data analytics, this paper proposes a survey encompassing multiple GPU database systems. The focus will be on elucidating the underlying mechanisms employed to deliver results and key performance metrics, utilizing benchmarks such as SSB and TPCH. This undertaking aims to shed light on new avenues for research within the realm of GPU databases.
\end{abstract}


\section{Introduction}
For decades, microprocessor performance consistently grew due to advancements in transistor speed and energy scaling. However, the decline in transistor-speed growth and physical energy limitations have introduced new challenges, slowing this growth. In response, research has shifted towards leveraging large-scale parallelism, heterogeneous cores, and accelerators for better performance and energy efficiency. Active exploration of software and hardware collaboration aims to achieve efficient data orchestration and energy-proportional computing.\cite{borkar2011future}

More Recently, the data analytics landscape has significantly evolved, notably embracing heterogeneous architectures like Graphics Processing Units (GPUs), Compute Express Link (CXL), and Smart Network Interface Cards (SmartNICs) to boost performance. This study focuses primarily on GPUs. In-memory analytics, often limited by memory bandwidth, has benefited greatly from GPUs' enhanced bandwidth capacities, promoting their widespread adoption in analytical query processing. Modern graphics cards possess high processing power and substantial memory bandwidth, rendering them formidable platforms for data-intensive applications. They excel in executing massive calculations on data in parallel \cite{gautam2013gpu}\cite{bress2014exploring}, thereby providing significantly faster data processing speeds compared to conventional Central Processing Units (CPUs).

Despite the extensive integration of GPUs in database systems, a gap exists in comprehensive comparative analyses among different GPU systems. This paper addresses this gap by proposing an in-depth investigation into multiple GPU database systems. The study examines essential performance metrics like Query Execution Time and explores the underlying architectures of these systems. The primary objective of the paper is to survey diverse GPU database systems, providing valuable insights into their potential and identifying areas for improvement.

\section{Previous Works}

Most current open source systems ~\cite{HeavyDB, BlazingSQL, PG-Storm} utilize a hybrid approach when it comes to GPU databases in that they perform most of their computation on the CPU and offload selected computation onto the GPU. Other systems ~\cite{TQP} first transform the data into other formats, such as PyTorch tensors, and perform kernel operations on them, incurring a high overhead. Purely research systems ~\cite{shanbhag2020crystal} that are GPU only support a limited number of operations due to one primary reason: A limited amount of data can be stored on a GPU. Some previous solutions to challenge have been (1) using compression to fit more data on the GPU ~\cite{gpubitpacking}, (2) utilizing multiple GPUs ~\cite{multi-gpu}. 

Previous work ~\cite{ssb-eval} have looked into benchmarking such systems but have primarily utilized SSB schema ~\cite{SSB} as a benchmark.  However, the SSB schema is a simplified version of the general TPC-H\cite{TPC-H} schema which contains more tables and a wider range of operations in its queries. Additionally, while previous work focuses on metrics such as execution time, which are important, they often fail to capture metrics related to the deployment of such systems in a production environment. Finally, most metrics reported by these systems are based on a performance on a singular system and never address how well these translate to other hardware configurations.

The approach to conducting a comprehensive survey and performance comparison of GPU database systems involved a well-defined methodology. The initial step was the careful selection of GPU database systems with an emphasis on diversity in features and use cases. BlazingSQL\cite{BlazingSQL}, HeavyDB\cite{HeavyDB}, TQP\cite{TQP}, and Crystal\cite{shanbhag2020crystal} were identified as the most suitable candidates for this comparative study. Execution time was established as the primary performance metric in the subsequent critical phase.

To ensure a thorough assessment, existing benchmark studies from prior research were examined, covering various scale factors against different benchmarks. Subsequently, experiments were conducted to obtain benchmark results for the selected systems. Notably, Crystal presented a challenge as it lacked support for TPCH queries, prompting additional efforts to address this limitation. Moreover, to fill a gap in previous work, a CPU-only OLAP DBMS comparison was introduced using DuckDB\cite{duckdb} as the benchmark tool. Data preparation involved creating representative datasets with scale factors up to 16 for SSB and up to 8 for TPCH.

The benchmarking workloads were strategically designed to encompass a spectrum of query complexities and data access patterns, facilitating a holistic evaluation of system performance. In the final phase, the findings from the experiments were synthesized, shedding light on both quantitative and qualitative differences between the GPU database systems under investigation. This systematic methodology provided valuable insights into the strengths and potential areas of improvement for each system.

~\section{Survey of Database Systems}
Understanding various GPU database systems is essential for appreciating the diverse methods these systems employ to manage workloads and execute queries. This section delves into different widely-used GPU databases, with the objective of understanding the unique architectures underpinning their designs. We also compare the performances of these GPU databases against DuckDB, which serves as a CPU-based baseline for our experiments. DuckDB is an in-process SQL OLAP database management system designed to support analytical query workloads. The GPU databases discussed in this section include:

\begin{itemize}
  \item \textbf{DuckDB (CPU)}
  \item \textbf{BlazingSQL}
  \item \textbf{OmniSciDB}
  \item \textbf{Crystal+}
  \item \textbf{Tensor Query Processor (TQP)}
\end{itemize}

Each of these GPU databases represents a unique approach to leveraging GPU technology for efficient data processing. By delving into their architectures and functionalities, one can gain a comprehensive understanding of the landscape of GPU-accelerated database systems and their respective contributions to handling varied workloads and queries.

\subsection{DuckDB}
DuckDB\cite{duckdb} is an in-process SQL OLAP (Online Analytical Processing) database management system designed to support complex analytical query workloads efficiently. Unlike traditional database management systems that require a server-client architecture, DuckDB is embedded directly within applications, making it highly accessible and easy to integrate. This in-process nature allows DuckDB to deliver high performance by eliminating the overhead of inter-process communication. It is particularly well-suited for data science and analytical tasks where it can be embedded within data analysis pipelines, scripts, or other data-centric applications. DuckDB offers robust SQL support, enabling users to execute complex queries with ease and efficiency. Additionally, it is optimized for handling columnar storage formats, which are crucial for analytical workloads that involve large-scale data scanning and aggregation. We used DuckDB as a CPU-based benchmark to compare the performance of the various GPU databases discussed in the following survey.

\subsection{BlazingSQL} \label{sec:blazingsql}
\subsubsection{Introduction}

BlazingSQL~\cite{BlazingSQL} functions as a robust SQL interface tailored for cuDF, offering advanced features to enhance data science workflows and manage large-scale enterprise datasets. Its notable integration with the dask-cudf library, part of the broader RAPIDS project, enhances adaptability and resilience, making BlazingSQL adept at addressing diverse data science tasks.

The adaptability and versatility of BlazingSQL are evident in its support for various formats and frameworks, positioning it as a crucial component in the data science ecosystem. Providing an efficient pathway from raw data handling to advanced analytics and machine learning, BlazingSQL streamlines the entire data science workflow, offering a cohesive solution. The open accessibility of BlazingSQL's foundational code, released under the Apache 2.0 License, reflects a commitment to transparency and community collaboration, fostering an open environment for development and innovation.

\subsubsection{Architecture and Features}

\begin{figure}
  \centering
  \includegraphics[width=\linewidth]{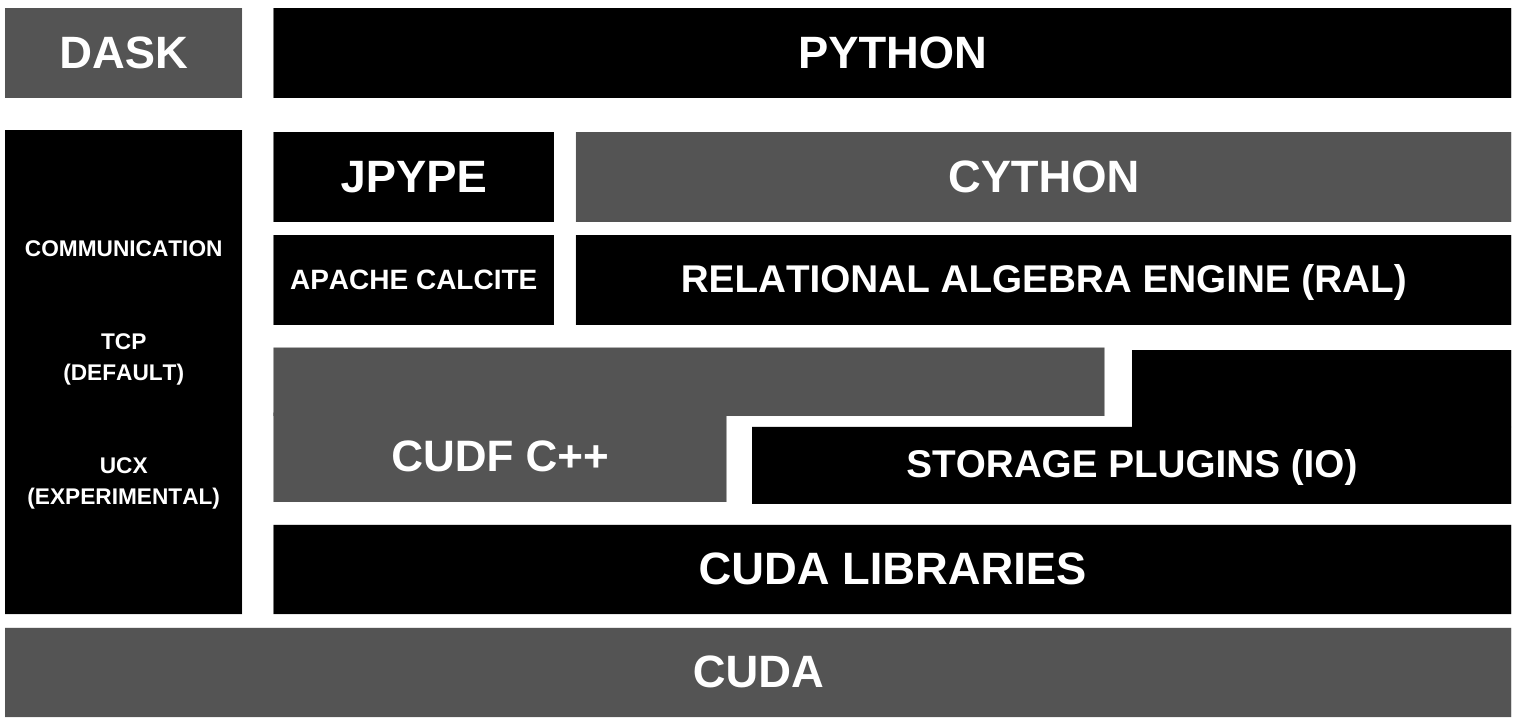}
  \caption{BlazingSQL RAL Architecture}
  \label{fig:blazingsql}
\end{figure}

Upon reviewing Figure \ref{fig:blazingsql}, it becomes evident that BlazingSQL initiates a connection with Apache Calcite via JPype. This connection serves a crucial role by utilizing Apache Calcite as a SQL parser, employing its capabilities to translate SQL strings into a relational algebra plan. The Relational Algebra Engine (RAL) plays a pivotal role in this intricate process, taking charge of generating a distributed homogeneous execution graph. This graph serves as a key instrument, communicating essential processing responsibilities to each worker within the system, thereby orchestrating an efficient and coordinated execution of the query.

The relational algebra in BlazingSQL transforms into a physical relational algebra plan, forming a directed acyclic graph (DAG) where nodes represent kernels and edges denote caches. Kernels logically organize transformations on distributed \texttt{DataFrames}, and \texttt{CacheData} objects, which can be in various states like \texttt{GPUCacheData} or \texttt{CPUCacheData}, facilitate data storage without immediate materialization. Most kernels utilize CacheData from a CacheMachine to create tasks for the task executor. The task executor, exclusive to Nvidia GPUs, manages resource access, handles memory, and supports retries for operations initially hindered by resource scarcity.

Upon residing as GPU \texttt{DataFrames} in GPU memory, users can leverage \texttt{RAPIDS cuML} for diverse machine learning applications or convert them to formats like \texttt{DLPack} or \texttt{NVTabular}, enabling in-GPU deep learning with frameworks like PyTorch or TensorFlow. In practical terms, BlazingSQL excels in Extract, Transform, Load (ETL) processes, efficiently transferring raw data straight into GPU memory, and transforming it into optimized GPU \texttt{DataFrames} for further analytical tasks.

BlazingSQL's proficiency extends to querying externally stored data, simplifying the process with standard SQL commands. The results manifest as \texttt{GPU DataFrames} (GDFs) in GPU memory, seamlessly integrating with RAPIDS libraries for diverse data science workloads. This capability enhances BlazingSQL's efficiency in integrating external data sources into GPU-accelerated analytics and processing pipelines

\subsubsection{Example of End-to-End Query Execution}
Consider the example given below:
\begin{lstlisting}[breaklines]{mysql}
Qx: SELECT o_custkey, SUM(o_totalprice) 
    FROM orders
    WHERE o_orderkey < 10
    GROUP BY o_custkey;
\end{lstlisting}

In the provided example query (\texttt{Qx}), when executed in the BlazingSQL Core engine, the SQL query undergoes parsing and optimization by Apache Calcite. The resulting optimized algebra, along with data sources like cudfs or files, is distributed to workers via \texttt{Dask}. The Relational Algebra (RAL) representation is as follows:

\begin{lstlisting}[breaklines]{ruby}
LogicalProject(o_custkey=[$0], 
EXPR$1=[CASE(=($2, 0), null:DOUBLE, $1)])
  LogicalAggregate(group=[{0}], EXPR$1=[$SUM0($1)], 
  agg#1=[COUNT($1)])
    LogicalProject(o_custkey=[$1], o_totalprice=[$2])
      BindableTableScan(
      table=[[main, orders]], filters=[[<($0, 10)]], 
      projects=[[0, 1, 3]], 
      aliases=[[$f0, o_custkey, o_totalprice]])
\end{lstlisting}

On each worker, the relational algebra is translated into a physical plan, exemplified below, where each relational algebra step corresponds to one or more physical plan steps. This physical plan constructs an execution graph forming a Directed Acyclic Graph (DAG) of kernels and caches. The cache's purpose is to store data as \texttt{CacheData} between computational stages, enabling data movement across different memory layers to scale beyond the capacity of a single layer.

\begin{lstlisting}[breaklines]{ruby}
LogicalProject(o_custkey=[$0],
EXPR$1=[CASE(=($2, 0), null:DOUBLE, $1)])
  MergeAggregate(group=[{0}], 
  EXPR$1=[$SUM0($1)], agg#1=[COUNT($1)])
    ComputeAggregate(group=[{0}], 
    EXPR$1=[$SUM0($1)], agg#1=[COUNT($1)])
      LogicalProject(o_custkey=[$1], o_totalprice=[$2])
        BindableTableScan(table=[[main, orders]], 
        filters=[[<($0, 10)]], projects=[[0, 1, 3]], 
        aliases=[[$f0, o_custkey, o_totalprice]])
\end{lstlisting}

Within the DAG, kernels are exclusively connected through caches. These kernels orchestrate complex distributed operations, generating tasks sent to the Task Executor for execution. The DAG's final output is a Cache containing the result.

\subsection{OmniSciDB} \label{sec:heavydb}

\subsubsection{Introduction}

OmniSciDB or HeavyDB is a state-of-the-art database that specializes in real-time analytics on large data sets through GPU acceleration. This cutting-edge platform is designed to optimize data processing, utilizing the raw power of GPUs to outperform traditional CPU-based databases. With its in-memory, columnar storage, and native SQL compatibility, HeavyDB ensures fast, efficient analytics operations.

A key feature of HeavyDB is its built-in visualization capabilities, enabling instant graphical data analysis and eliminating the need for separate visualization tools. Known for speed, scalability, and adaptability, HeavyDB caters to various sectors, including finance and telecommunications, where rapid data analysis is essential.

Despite its reliance on specific hardware, HeavyDB's ability to integrate with common data science environments and scale with additional GPU resources makes it a versatile and powerful tool in the realm of big data and analytics. As a leader in GPU-accelerated data processing, HeavyDB is pivotal for businesses aiming to achieve swift and insightful data-driven decisions.

\subsubsection{Architecture}
\begin{figure}
  \centering
  \includegraphics[width=\linewidth]{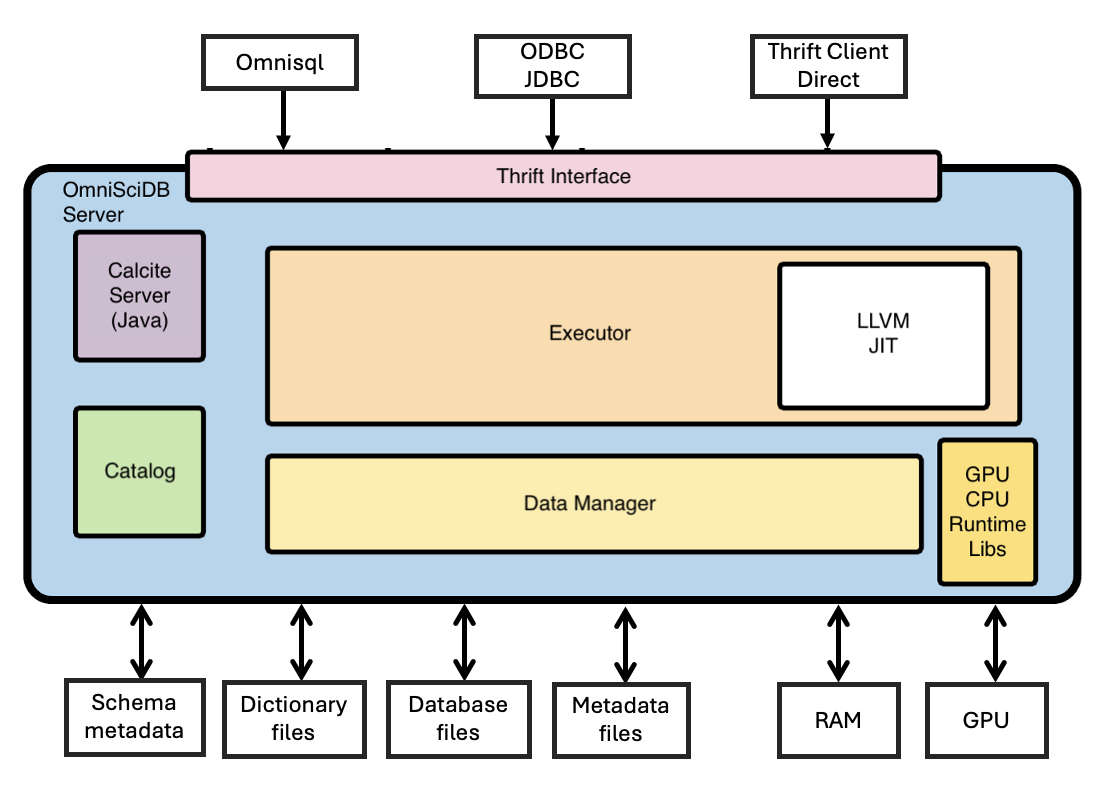}
  \caption{OmniSciDB High-Level Architecture}
  \label{fig:heavydb}
\end{figure}

OmniSciDB distinguishes itself with a GPU-centric processing approach, ensuring each operation is finely tuned for GPU efficiency. At the heart of its architecture as shown in Figure \ref{fig:heavydb}, lies Apache Thrift, which standardizes communication for both external clients and internal processes, facilitating seamless interactions with various client tools, including the command line interface omnisql, JDBC driver, and \texttt{SQLImporter} utility.

For query optimization, it leverages Apache Calcite, renowned for its modular nature and flexibility. This enables the addition of custom functions to the SQL parsing process, such as trigonometric computations vital for geospatial analysis, ensuring seamless integration and optimization of these functions within the query plans.

The database's structure is further strengthened by its Catalog component, which manages metadata through a centralized system. Each database maintains its own Catalog, all orchestrated under a System Catalog that actively manages metadata repositories via an SQLite database. This centralized metadata management exemplifies OmniSciDB's layered approach to organizing and retrieving data efficiently.

Adding to these features, it incorporates an integrated visualization engine, directly transforming data into visual analytics and enhancing the data-to-insight transition. Its robust SQL interface provides a familiar landscape for traditional database users, making the shift to this powerful GPU-driven analytics platform smoother. The system's advanced memory management and cross-filtering functionalities significantly bolster its data processing capabilities, while its columnar data storage format ensures fast read operations and optimal data compression.

Through this integration of sophisticated components, it emerges as a highly efficient and scalable solution, capable of handling diverse analytical workloads and leading the way in the realm of GPU-accelerated data analytics.

\subsubsection{Example of End-to-End Query Execution }

Consider the following example query:

\begin{lstlisting}[breaklines]{mysql}
SELECT o_custkey, SUM(o_totalprice) 
FROM orders
WHERE o_orderkey < 10
GROUP BY o_custkey;
\end{lstlisting}

When executed in the OmniSci (HeavyDB) engine, this SQL query undergoes a series of steps designed to ensure high-performance execution. The execution process begins with parsing and validating the SQL query to ensure its syntactic correctness. Once validated, the query is converted into an optimized sequence of relational algebra operations by the OmniSci planner. This optimized logical plan ensures efficient query execution by leveraging the database's capabilities.

The logical plan for this query can be represented as follows:

\begin{lstlisting}[breaklines]{ruby}
LogicalProject(o_custkey=[$0], total_price=[$1])
  LogicalAggregate(group=[{0}], total_price=[$SUM0($1)])
    LogicalFilter(condition=[<($2, 10)])
      LogicalProject(o_custkey=[$1], o_totalprice=[$2], o_orderkey=[$0])
        BindableTableScan(table=[[main, orders]], projects=[[o_orderkey, o_custkey, o_totalprice]])
\end{lstlisting}

Following the generation of the optimized relational algebra sequence, the execution environment is prepared. This involves setting up the necessary resources and configurations required for executing the query efficiently. In the subsequent step, data ownership and identification processes ensure that the necessary data is accessible. The relevant rows from the \texttt{orders} table, where \texttt{o\_orderkey} is less than 10, are identified and loaded onto the target devices (CPU/GPU) as required.

Once the data is prepared, the query kernel is executed on the target devices. This step involves applying the filtering operation to select rows with \texttt{o\_orderkey < 10}, grouping the selected rows by \texttt{o\_custkey}, and performing the \texttt{SUM(o\_totalprice)} aggregation for each group. The GPU acceleration capabilities of OmniSci are leveraged during this step to achieve high-speed computation.

After the query kernel execution, the partial results from different execution threads and devices are combined in the result reduction phase. This step ensures that the final sum for each customer key is accurately computed. A check is then performed to determine if the query execution is complete. If additional steps are required, they are executed accordingly.

The final step involves formatting the result set according to the query requirements and returning it to the client. The physical plan representation for the execution steps is as follows:

\begin{lstlisting}[breaklines]{ruby}
Project(o_custkey=[$0], total_price=[$1])
  Aggregate(group=[{0}], total_price=[$SUM0($1)])
    Filter(condition=[<($2, 10)])
      Project(o_custkey=[$1], o_totalprice=[$2], o_orderkey=[$0])
        TableScan(table=[[main, orders]], projects=[[o_orderkey, o_custkey, o_totalprice]])
\end{lstlisting}

This detailed flow ensures that the query execution in OmniSci DB aligns with the official process, providing high-performance and accurate results through the use of optimized relational algebra sequences, efficient data handling, and GPU acceleration.

\subsubsection{Advantages and Industry Application}
The real-time processing and visualization capabilities of HeavyDB make it invaluable across various sectors, including financial services, telecommunications, retail, and public safety. Its GPU-first approach provides speed and performance, with scalability achieved through horizontal GPU integration. Although requiring specific GPU hardware presents a limitation, HeavyDB's ability to integrate with data science tools like Python and R compensates by adding versatility. As data volumes grow and the demand for quick analytics increases, HeavyDB's focus on GPU utilization and real-time data handling solidifies its position as a key player in data management and analysis.

\subsection{Crystal+} \label{sec:crystal}

\subsubsection{Introduction}
Crystal+ is a collection of block-wide device functions that can be used to implement high-level SQL queries based on the original Crystal library ~\cite{shanbhag2020study}. In this section, we will cover the original Crystal library and then dive into the improvements made in the next section. The Crystal library stores the table in columnar format as well as perform dictionary encoding ~\cite{dictionary_encoding} to convert all non-numerical data to a numerical format. Crystal stores the working set of the data we are operating on in the GPU memory itself rather than using a coprocessing model to move the data to the GPU during query time. The benefit that Crytal has over the other data in GPU solutions is that it uses a tile-based execution model. Specifically, the elements we want to work on are broken into tiles and each thread block is responsible for one tile. This also involves loading the data into shared memory of the GPU once and then performing all operations in shared memory thus limiting the I/O bandwidth to fetch and write the data. It then implements a series of device functions such as \texttt{BlockAggergate} that utilize the tile-based execution model and then combine these blocks to implement various SSB queries. Note that these atomics are relatively limited and can't be combined to implement all queries, such as the TPC-H benchmark. Additionally, all of the Crystal implementation relies on internal data structures and doesn't take advantage of the advancements made by NVIDIA in the CUDA programming framework. 

\subsubsection{Improvements made}

\begin{figure}[h]
\centering
\includegraphics[scale=0.425]{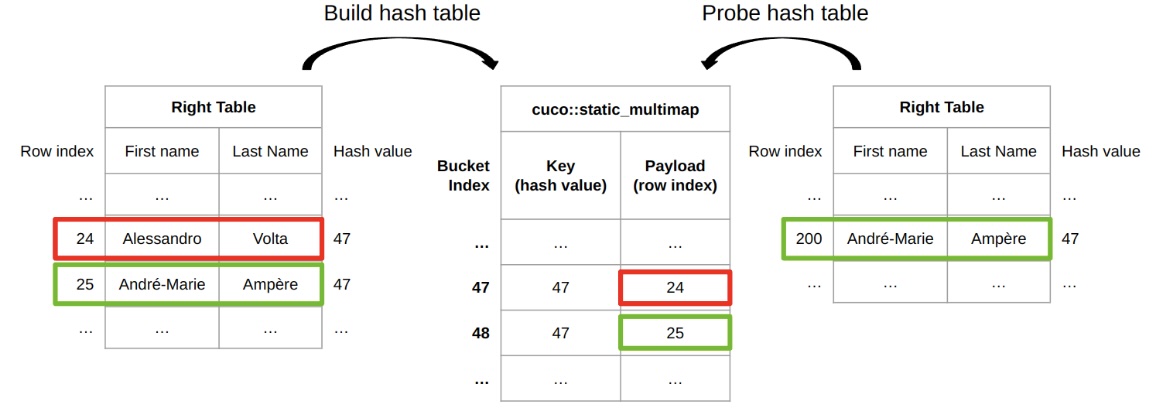}
\caption{How Crystal+ implements hash-based join}
\label{fig:hash_join}
\end{figure}

As mentioned above, the original Crystal doesn't take advantage of the advancements in GPU programming. In order to do so, the original blocks implemented in the Crystal repo were reimplemented using the Thrust ~\cite{Thrust}, Cub ~\cite{Cub}, and cuCollections ~\cite{cuCollections} libraries. Not only does this give better performance than the naive array-based implementations as we increase the size of the dataset but also is more modular and adaptable to take advantage of the improvements made in different NVIDIA hardware. 

One of the main operations that is performed in most analytics SQL queries is joins and we can take advantage of the cuCollections library to implement an efficient hash-based join as illustrated in Figure ~\ref{fig:hash_join}. This involves two different phases of build phase and probe hash similar to the implementation of hash-based join on the CPU. However, we are going to be using cuCollection's multi-map to store the data and thus each thread in the kernel in the build and probe phase just needs to be responsible for determining the row's key based on the row's values. 

One another common operation that is performed in many queries is multi-column group by and thus Crystal+ implements an efficient group by using the static map. It does so by determining the maximum value for each column and using that to compute the number of bits necessary to represent all of the values in the column which we will call width. It then uses the width to determine a bitmask as well as a key offset for each column. Then for each row, it first applies the bit mask for each of the column's values and then performs the bit shift for that value and then combines the values to generate a group key. It then performs the associated aggregation for that group and each thread does it for the row it is responsible for. Once the kernel has finished execution, we can get the aggregate for each group as well as use the bit mask and bit shift to extract the values that made up the group.   

\subsubsection{Query Execution}

In order to determine the block functions we need to combine for a given query, we utilize Postgres's \texttt{EXPLAIN} ~\cite{postgres_explain} functionality to generate a query plan. Consider the example query:

\begin{lstlisting}[breaklines]{mysql}
    SELECT l_returnflag, l_linestatus, COUNT(*),
    FROM lineitem, orders,
    WHERE lineitem.orderkey = orders.orderkey,
    GROUP BY l_returnflag, l_linestatus;
\end{lstlisting}

Running \texttt{EXPLAIN} in Postgres for this query gives us a query plan of:
\begin{align*}
&\text{Finalize GroupAggregate}  \\
&\quad \text{Group Key: lineitem.l\_returnflag, lineitem.l\_linestatus} \\
&\quad \rightarrow  \text{Gather Merge}  \\
&\quad \quad \text{Workers Planned: 2} \\
&\quad \quad \rightarrow  \text{Sort}  \\
&\quad \quad \quad \text{Sort Key: lineitem.l\_returnflag, lineitem.l\_linestatus} \\
&\quad \quad \quad \rightarrow  \text{Partial HashAggregate}  \\
&\quad \quad \quad \quad \text{Group Key: lineitem.l\_returnflag, lineitem.l\_linestatus} \\
&\quad \quad \quad \quad \rightarrow  \text{Parallel Hash Join} \\
&\quad \quad \quad \quad \quad \text{Hash Cond: (lineitem.l\_orderkey = orders.o\_orderkey)} \\
&\quad \quad \quad \quad \quad \rightarrow  \text{Parallel Seq Scan on lineitem}  \\
&\quad \quad \quad \quad \quad \rightarrow  \text{Parallel Hash} \\
&\quad \quad \quad \quad \quad \quad \rightarrow  \text{Parallel Scan using orders} \\
\end{align*}

Anytime the query plan mentions a join, we utilize the hash joined mentioned the above section and when the query plan mentions an aggregate we utilize the aggregate mentioned in the section above. 

\subsection{Tensor Query Processor (TQP)} \label{sec:tqp}

\subsubsection{Introduction}
Tensor Query Processor (TQP) \cite{TQP} represents a novel paradigm in analytical database management systems, aiming to leverage the strengths of both relational databases and tensor computing within a unified framework. TQP transforms traditional SQL queries into tensor programs and executes them on TCRs such as PyTorch, TensorFlow, TVM, ONNX, etc. TQP introduces a set of novel algorithms and a compiler stack for converting relational operators into tensor computations. This helps TQP achieve the following three goals:
\begin{itemize}
    \item \textbf{Performance:} Deliver significant performance improvements over CPU-based data systems, and match or outperform custom-built solutions for GPUs. TQP capitalizes on the computational prowess of TCRs and provides a TCR-aware query optimizer to extract the best performance from the underlying hardware.
    \item \textbf{Portability:} Demonstrate portability across a wide range of target hardware and software platforms. Being hardware agnostic allows TQP to adapt its execution strategies to ensure optimal performance across diverse computing environments ranging from CPUs, to discrete GPUs, integrated GPUs (Intel and AMD), NN-accelerators (TPUs), and web browsers.
    \item \textbf{Parsimony:} Prioritize developer productivity by providing a robust, flexible and sustainable framework. TQP's sophisticated compiler stack automates the translation of SQL to optimized tensor programs, reducing the need for manual coding and intricate optimization techniques. 
\end{itemize}

\subsubsection{Architecture and Features}

\begin{figure}
  \centering
  \includegraphics[width=1.0\linewidth]{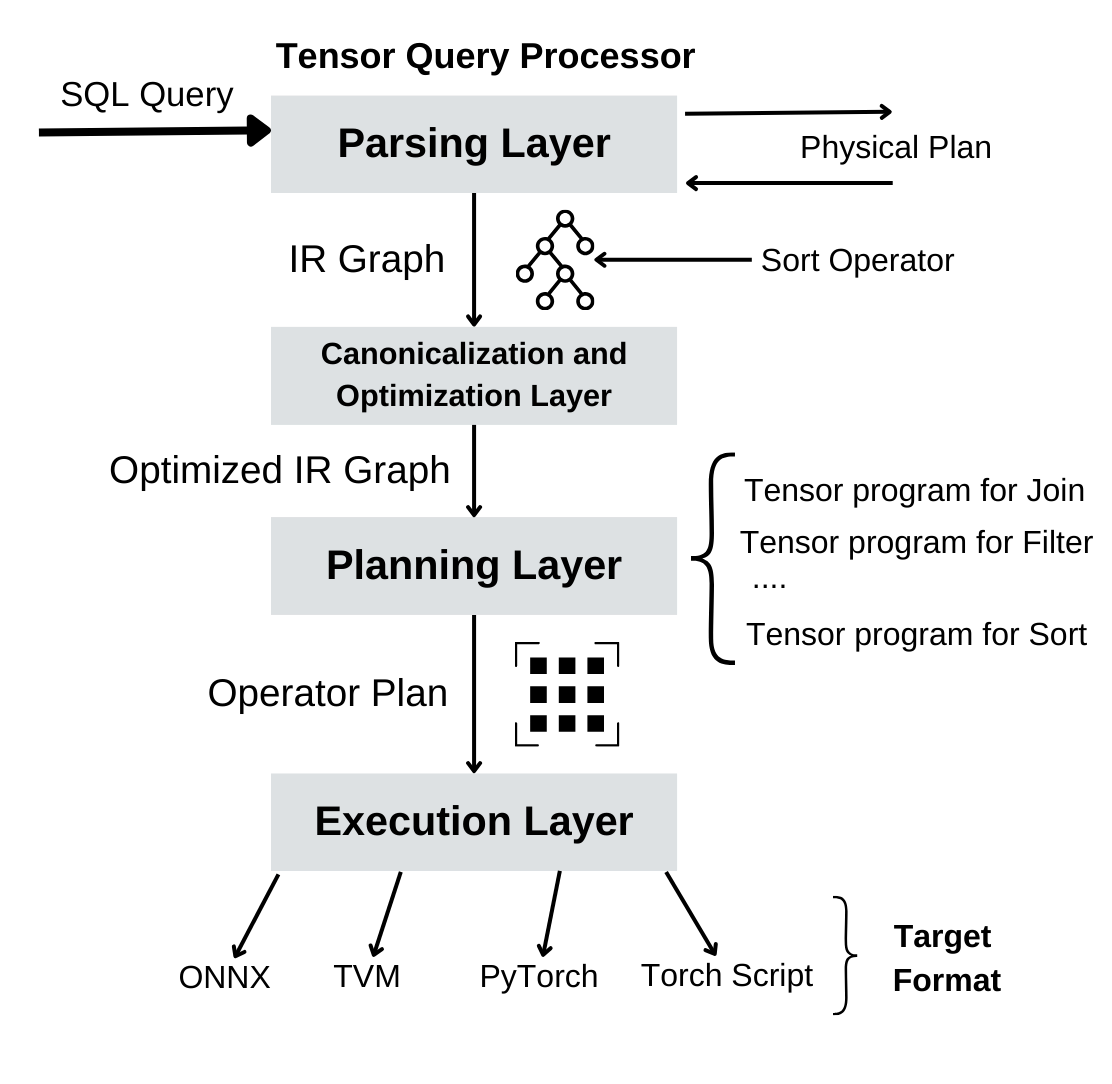}
  \caption{TQP's compilation phase}
  \label{fig:tqp}
\end{figure}

\begin{figure}
  \centering
  \includegraphics[width=1.0\linewidth]{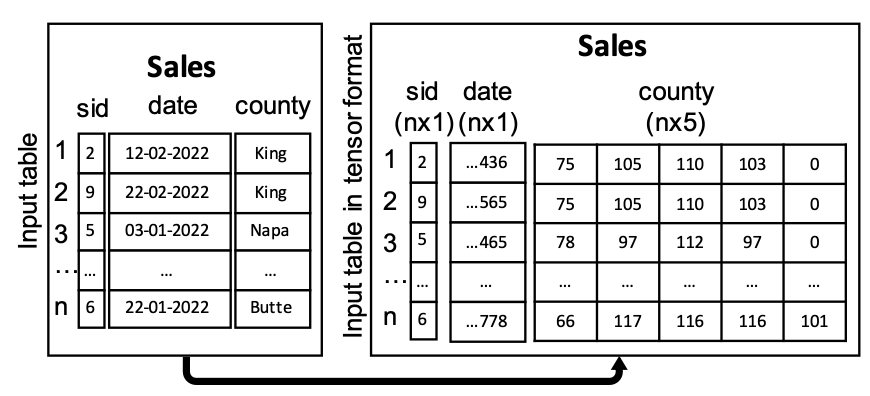}
  \caption{TQP's represents data in tensors}
  \label{fig:tqp_data}
\end{figure}

TQP operates in two phases: 
\begin{enumerate}
    \item \textbf{Compilation:} Figure \ref{fig:tqp} illustrates the four-step transformation of an SQL query into an executable tensor program. 
    \begin{itemize}
        \item Parsing Layer: Constructs an internal \textit{intermediate representation (IR)} graph depicting the query's physical plan. 
        \item Optimization Layer: Performs canonicalization and applies optimization rules to yield an \textit{optimized IR graph}.
        \item Planning Layer: Translates the optimized IR graph into an \textit{operator plan} containing a mapping of each operator to a tensor program implementation.
        \item Execution Layer: Generates an \textit{executor} object responsible for orchestrating the tensor program execution. It sequentially processes data tensors while managing memory through garbage collection. Additionally, it supports dynamic compilation for various target formats (e.g., PyTorch, TVM, ONNX).
    \end{itemize}
    
    \item \textbf{Execution:} This phase involves transforming the input data into tensors (refer Figure \ref{fig:tqp_data} and feeding it into the executor object, returning the query result in tensor, NumPy, or Pandas format. TQP exploits the tensor-level intra-operator parallelism provided by the TCRs.
\end{enumerate}

The code sample below demonstrates submitting a query string to the TQP compiler and then running the compiled query object generating the output in Pandas dataframe format.

\begin{lstlisting}[breaklines]{bash}
statement = \
    "SELECT Digits, Sizes, COUNT(*)
    FROM numbers GROUP BY Digits, Sizes"
    
compiled_query = \
    tqp.sql.query.spark(statement, device="cuda")
    
compiled_query.run(toPandas=True)
\end{lstlisting}

Following is an overview of the supported operators in TQP:
\begin{enumerate}
    \item Relational operators such as selection, projection, sort, group-by aggregation (sort-based), natural join (hash and sort-based), non-equi, left-out, left-semi and left-anti joins.
    \item Comparison and arithmetic operators, date functions and Nulls.
    \item Statements such as \texttt{IN}, \texttt{CASE} and \texttt{LIKE} clauses.
    \item Aggregates like \texttt{SUM, AVG, MIN, MAX, COUNT, DISTINCT COUNT}.
    \item Scalar, Nested and Correlated subqueries.
    \item Prediction Queries: TQP provides seamless integration with PyTorch models and traditional ML models using Hummingbird \cite{Hummingbird}, enabling native support for predictive capabilities. A Prediction Query, which encapsulates a trained ML model making predictions on the input data, can feature a combination of ML operations such as tree ensemble, one-hot encoding, scaling, and concatenation. These operations may be complemented by relational operators like join, aggregation, and filtering to create a comprehensive predictive model.
\end{enumerate}

\section{Evaluation}
\begin{figure*}[t]
    \begin{subfigure}{\textwidth}
        \centering
        \includegraphics[width=\linewidth]{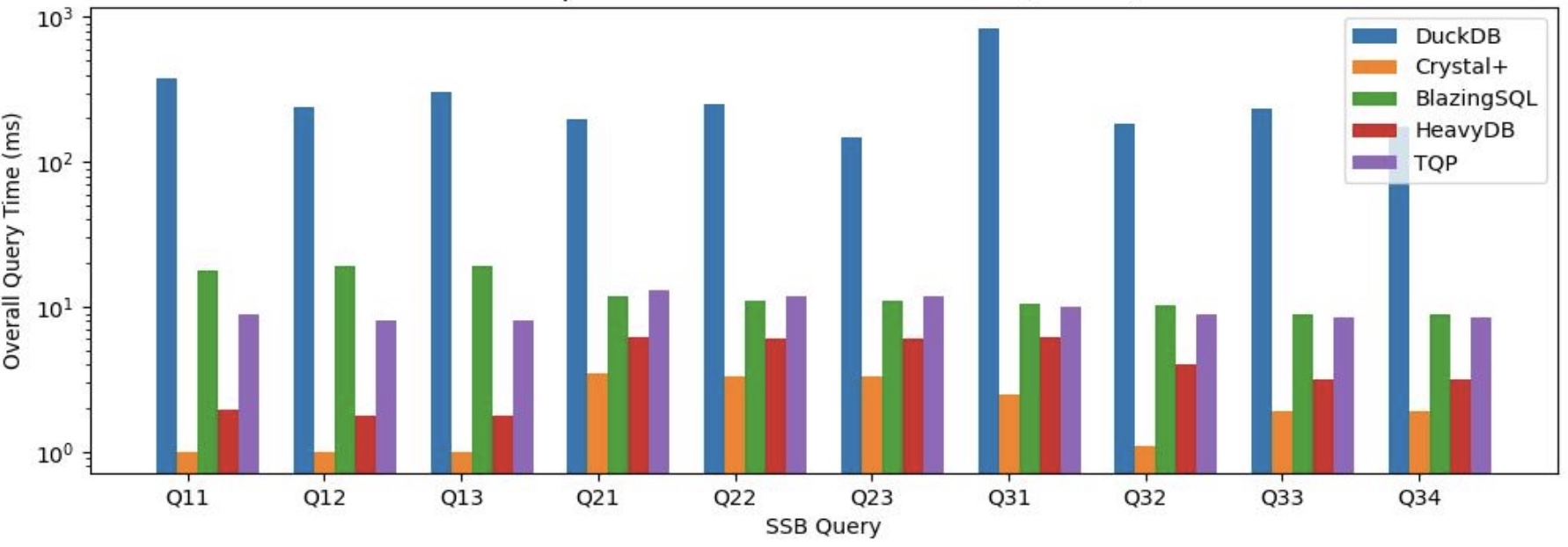}
        \caption{Comparison on the SSB benchmark (SF = 16)}
        \label{fig:ssb_comparison}
    \end{subfigure}
    
    \begin{subfigure}{\textwidth}
        \centering
        \includegraphics[width=\linewidth]{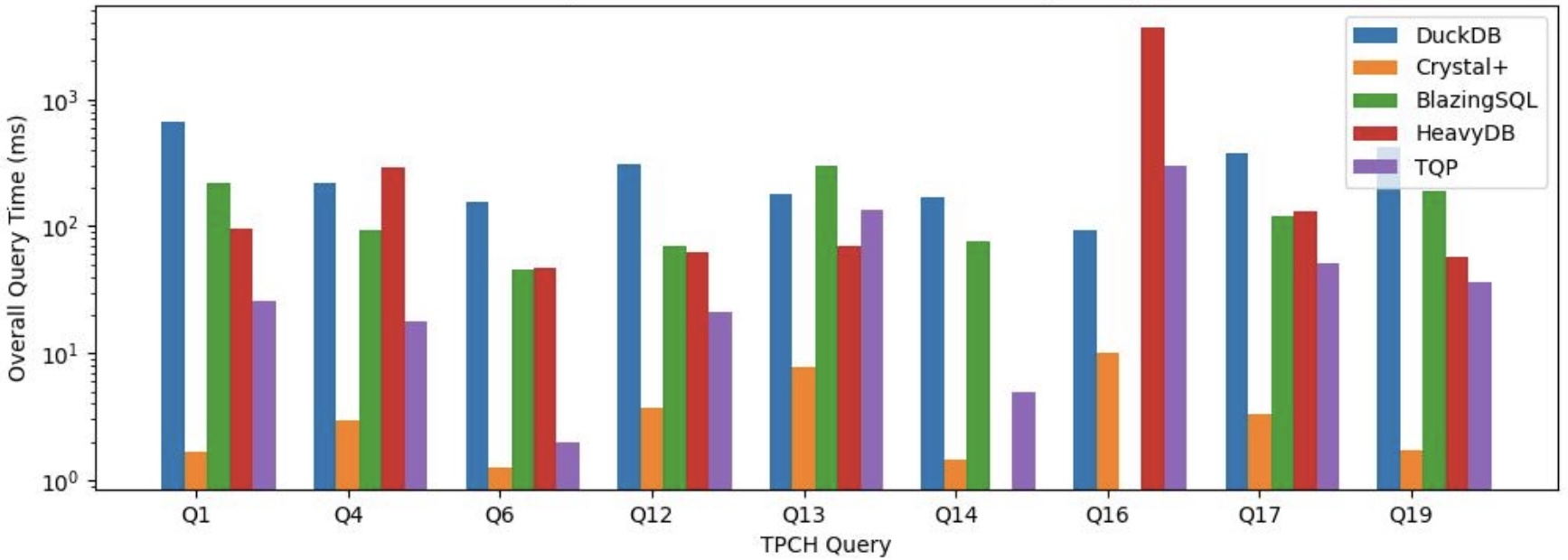}
        \caption{Comparison on the TPCH benchmark (SF = 1)}
        \label{fig:tpch_comparison}
    \end{subfigure}

    \caption{Overall runtime of various databases on many benchmark queries}
    \label{fig:overall_comparison}
\end{figure*}

\subsection{Quantitative Comparison}

We benchmarked these databases on both the SSB and TPCH benchmarks and compared the performance of these databases with each other as well as DuckDB, which is a common OLAP CPU-based database. The performance of these databases on different SSB and TPCH queries can be found in Figure ~\ref{fig:overall_comparison}. For the TPCH benchmark, we were originally planning to benchmark it using SF = 8 but we could only find data for TQP, which is closed source, for SF = 1. 

Looking at the performance on the SSB (SF = 16) queries, we see that DuckDB overall takes the most time followed by BlazingSQL and TQP taking the same amount of time, HeavyDB generally performs a little better than them with Crystal+ outperforming all of these systems substantially. Looking at Q11, Q12, and Q13 queries we see that BlazingSQL takes around 18.5 ms while TQP takes around 8.5 ms even though they perform equally on the other queries. For Q21, Q22, and Q23 we see all of the GPU databases have equal runtimes with limited variances but DuckDB runtimes range from 150 to 250 ms. For Q31, Q32, Q33, and Q34 we see that Crystal+ takes only 1.1 ms for Q32, takes around 1.9 ms for Q33 and Q34 and takes 2.5 ms for Q31.

Looking at the TPCH (SF = 1) queries, there are a couple of things that stand out. Note that we used SF = 1 because of the fact that TQP only reported numbers for SF = 1 and since it is a closed source we can't benchmark its performance on other queries. We see that for Q4, DuckDB takes 216 ms while HeavyDB takes 292 ms and for Q13 DuckDB takes 181 ms while BlazingSQL takes 303 ms. This is especially visible in Q16 as DuckDB takes 93 ms, HeavyDB takes 3689 ms and TQP takes 301 ms. This is interesting that some of the leading GPU databases actually take more time than DuckDB, a common CPU-based OLAP solution for complicated queries as TPCH queries are generally considered more challenging than SSB. 

One thing that is common across both of these benchmarks is that Crystal+ outperforms all of these databases. We see on the SSB benchmark that HeavyDB takes on average 4.052 ms while Crystal on average takes 2.05 ms which is a 1.97x improvement. This difference is especially pronounced in the TPCH benchmark where TQP takes 66.22 ms while Crystal+ takes 3.75 ms, which is a 17.66x improvement. 

\subsection{Qualitative Comparison}

\begin{table*}[t]
    \centering
    \begin{tabular}{|C{2cm}|C{3cm}|C{3cm}|C{3cm}|C{3cm}|}
        \hline
        \textbf{Feature/Aspect} & \textbf{BlazingSQL} & \textbf{HeavyDB (OmniSciDB)} & \textbf{CrystalDB} & \textbf{Tensor Query Processor (TQP)} \\
        \hline
        \textbf{Underlying Tech} & RAPIDS Ecosystem, Apache Calcite, cuDF & GPU acceleration, SQL engine, geospatial support & Serverless PostgreSQL & Tensor Computation Runtimes (e.g., PyTorch, TVM) \\
        \hline
        \textbf{Data Processing} & Using GPU DataFrames & High performance, vectorization, advanced memory mgmt & Auto-scaling, optimized for speed and reliability & Columnar Tensor-based data format \\
        \hline
        \textbf{Performance} & Optimized Query Compilation, In-Memory operations and GPU acceleration & Rapid query compilation, native SQL support & Managed configuration for performance optimization & High performance over CPU, comparable to GPU systems \\
        \hline
        \textbf{Special Features} & Interoperability with RAPIDS libraries and External Data Source Integration & Server-side rendering, web-based visualization (licensed) & Multi-tenancy management, security, compliance & Unified runtime for relational and ML operators; hardware agnostic \\
        \hline
        \textbf{Use Case} & Large-Scale Data Analytics and Machine Learning with cuML Integration & Analytical platforms, geospatial data & SaaS providers, multi-cloud deployments & ML assisted Analytical workloads \\
        \hline
        \textbf{Development Effort} & SQL Compatibility and Integration with Existing Ecosystems & Requires a license for full feature set & DBA optional, self-managing & Minimal engineering effort, Closed-source and in development phase. \\
        \hline
    \end{tabular}
    \caption{Comparative Architectural Features}
    \label{table:comparative_features}
\end{table*}

A qualitative comparison of these systems can be found in Table ~\ref{table:comparative_features}. The comparative analysis reveals distinctive characteristics among the considered database systems. BlazingSQL, deeply integrated with the RAPIDS ecosystem and cuDF, optimizes query compilation through GPU acceleration and in-memory operations. It particularly excels in large-scale data analytics and machine learning, emphasizing interoperability with RAPIDS libraries. HeavyDB, or OmniSciDB, stands out for analytical platforms and geospatial data support, leveraging GPU acceleration for high performance and featuring server-side rendering. CrystalDB targets SaaS providers with its multi-tenancy management, security, and compliance features, making it suitable for multi-cloud deployments. TQP uniquely focuses on analytical DBMSs and AI workloads, employing SQL-to-tensor program transformations and machine-level code compilation for efficient tensor computation.

In terms of data processing, BlazingSQL utilizes \texttt{GPU DataFrames}, while HeavyDB emphasizes vectorization and advanced memory management. CrystalDB boasts auto-scaling and optimization for speed and reliability, while TQP integrates SQL-to-tensor program transformations and machine learning. The performance of these systems is context-dependent, with BlazingSQL showcasing optimized query compilation and HeavyDB featuring rapid query compilation and native SQL support. CrystalDB excels in managed configuration for performance optimization, while TQP demonstrates high performance comparable to GPU systems over CPU. Each system brings special features to the table, such as BlazingSQL's interoperability with RAPIDS libraries and external data source integration, HeavyDB's server-side rendering and licensed web-based visualization, CrystalDB's emphasis on multi-tenancy and security, and TQP's support for relational operations and machine-level code compilation. The development effort varies, with BlazingSQL prioritizing SQL compatibility and ecosystem integration, HeavyDB requiring a license for its full feature set, CrystalDB being designed for self-management with optional DBA involvement, and TQP featuring parsimonious engineering effort with portability across hardware. These distinctions underline the nuanced suitability of each system for specific use cases and deployment scenarios.

Note that we only compare complete GPU systems and thus don't include Crystal because it is a GPU library rather than a complete GPU database with features such as query plan. Additionally, these comparisons are based on the opinions of the author and are more subjective. 

\section{Conclusion}

This paper provides an in-depth overview of four distinct systems designed to enhance database performance through GPU acceleration: BlazingSQL, OmniSciDB, Tensor Query Processor (TQP), and Crystal+. Each section explores the features, implementation details, and GPU utilization in executing queries for these systems, illustrated with examples. The paper also highlights the distinct features and use cases of each of the database systems mentioned. BlazingSQL leverages the RAPIDS ecosystem and cuDF for data processing, focusing on GPU DataFrames. HeavyDB emphasizes GPU acceleration and native SQL support, particularly for geospatial data. CrystalDB operates as a serverless PostgreSQL with auto-scaling capabilities and emphasizes reliability. TQP integrates tensor computation runtimes like PyTorch and TVM, supporting tensor-based data formats and unified relational and ML operations. Each system varies significantly in its underlying technologies, data processing methods, performance optimizations, special features, use cases, and developmental considerations, highlighting their unique approaches to GPU-accelerated database management. Performance comparisons on TPCH and SSB benchmarks, along with a qualitative assessment, are included.

To conduct a more comprehensive quantitative analysis, open-sourcing all systems is imperative, although currently unavailable. The goal is to delve into metrics like GPU/CPU utilization, IO operations, and performance across diverse benchmarks and sizes. Further exploration encompasses aspects such as reliability, transaction guarantees, security, privacy, scalability, and simulating production-level workloads to evaluate the associated costs. As GPU databases are in their early stages, ongoing exploration aims to understand limitations hindering production deployment and strategize solutions to overcome these challenges. Through such comprehensive analysis and exploration, the potential for advancements in GPU database systems can be fully realized, paving the way for future innovations in this evolving field.

\section*{Acknowledgment}

We would like to express our sincere gratitude to the University of Wisconsin-Madison for their invaluable support and resources that facilitated this survey project. Our deepest thanks go to Professor Xiangyao Yu for his expert guidance, insightful feedback, and unwavering encouragement throughout the duration of this study. His expertise and mentorship were instrumental in shaping the direction and quality of our research. We are also grateful to the faculty and staff who provided additional assistance and resources, making it possible to complete this project successfully.

\end{document}